\documentclass{emulateapj}

\begin{document}

\title{Monte Carlo radiation transfer simulations of photospheric
  emission in long-duration Gamma-Ray Bursts}

\shorttitle{MC Radiation Transfer in GRBs}

\author{Davide Lazzati\altaffilmark{1}}

\shortauthors{Lazzati}

\altaffiltext{1}{Department of Physics, Oregon State University, 301
Weniger Hall, Corvallis, OR 97331, U.S.A.}

\begin{abstract}
  We present MCRaT, a Monte Carlo Radiation Transfer code for
  self-consistently computing the light curves and spectra of the
  photospheric emission from relativistic, unmagnetized jets. We apply
  MCRaT to a relativistic hydrodynamic simulation of a long duration
  gamma-ray burst jet, and present the resulting light-curves and
  time-dependent spectra for observers at various angles from the jet
  axis. We compare our results to observational results and find that
  photospheric emission is a viable model to explain the prompt phase
  of long-duration gamma-ray bursts at the peak frequency and above,
  but faces challenges in reproducing the flat spectrum below the peak
  frequency. We finally discuss possible limitations of these results
  both in terms of the hydrodynamics and the radiation transfer and
  how these limitations could affect the conclusions that we present.
\end{abstract}

\keywords{gamma-ray burst: general --- radiation mechanisms: thermal
  --- radiative transfer }

\section{Introduction}

Photospheric emission is a leading model to explain the prompt
radiation of long-duration gamma-ray bursts
\citep{Rees05,Giannios06,Lazzati09,Lundman13,Peer15}.  In the
photospheric model the radiation is assumed to form deep in the
outflow, where the jet is highly optically thick, and to be advected
out, eventually being released when the jet becomes transparent. The
photospheric model does not specify the radiation mechanism involved,
since the spectrum is shaped by the interaction with matter, mostly
through Compton scattering, and not by the emission process itself.

The alternative model, the Synchrotron Shock Model (SSM,
\citealt{Rees94,Daigne98,Lloyd02,Daigne11,Zhang11}), assumes instead
that the jet crosses the photosphere virtually deprived of radiation,
coasts until the Thomson optical depth is well below unity, and is
reactivated when shells of different speed shock each other and
generate non-thermal particles and magnetic field \citep{Rees94}. SSM
predicts that the radiation spectrum is directly related to the
radiation mechanism and suffers little interaction with the baryons
and leptons of the jet once it is produced. Both models have been
successful in explaining some of the observed properties of long
duration GRBs but are in significant tension with other
properties. For example, SSM easily accounts for the non-thermal,
broadband nature of the prompt spectrum \citep{Piran99}, but faces
challenges for its intrinsic width \citep{Vurm16} and when compared to
ensemble correlations such as the Amati, Yonetoku, Golenetskii, or
E-$\Gamma$ correlations
\citep{Golenetskii83,Amati02,Yonetoku04,Liang10,Ghirlanda12}. On the
other hand, the photospheric emission has been shown to be able to
reproduce all the correlations (\citealt{Lazzati11,Lazzati13},
hereafter L13; \citealt{Lopez14}) but can hardly produce a spectrum
that has prominent non-thermal tails at both low- and
high-frequencies. While sub-photospheric dissipation can cure the
high-frequency problem producing prominent non-thermal tails
\citep{Peer06,Giannios06,Giannios07,Beloborodov10,Lazzati10,Vurm11,Ito13,Ito14,Chhotray15,Santana16},
it is still unclear whether the low-frequency tails can be reproduced
in photon-starved conditions without invoking an external radiation
mechanism such as synchrotron \citep{Peer11,Chhotray15}. Encouraging
results, however, have been obtained for photospheric emission with
contributions from an subdominant synchrotron component
\citep{Vurm16}.

In both cases, theoretical models only partially address the
complexity of the GRB phenomenon. With very few exceptions, models
either simplify the radiation physics or the geometry of the emitting
region.  For example, photospheric radiation computed from jet
simulations -- and therefore fully taking into account the complexity
of structured and turbulent outflows -- assume that the
radiation-matter coupling is infinite below the photosphere but
suddenly drops to zero as soon as the photosphere is crossed (e.g.,
\citealt{Lazzati09,Mizuta11,Nagakura11}; L13). On the other hand,
models that consider proper radiation transfer or radiation from
magnetized jets with dissipation and particle acceleration are applied
to simplified homogeneous plasma regions or to spherical jets with
analytic radial stratification and without any polar structure (e.g.,
\citealt{Peer06,Giannios07,Beloborodov10,Lazzati10,Vurm11,Chhotray15,Vurm16,Beloborodov16}).

In this paper we present the results of a first step towards merging
the two worlds and performing detailed radiation transfer calculations
on the complex dynamics of a relativistic jet from high-resolution
multi-dimensional numerical simulations (analogous to the recent work
by \citealt{Ito15}). In Section~\ref{sec:mcrat} we present the Monte
Carlo Radiation Transfer (MCRaT) code and the results of several
validation tests. In Section~\ref{sec:results} we present and discuss
the results of applying the code to a hydrodynamic (HD) simulation of
a GRB jet. Finally, in Section~\ref{sec:discussion} we discuss our
results by comparing them with observations and previous theoretical
work, and we discuss limitations and possible improvements of this
technique.

\section{The MCRaT code}
\label{sec:mcrat}

We developed a Monte Carlo Radiation Transfer (MCRaT code) to compute
time-resolved light curves and spectra from hydrodynamic simulations
of GRB jets. The approach to the single-photon scattering in the code
is to neglect the energy losses of the electron population, randomly
selecting at each scattering an electron from a thermal population at
a temperature that does not depend on previous scatterings. This is a
fairly standard approach that assumes the heat capacity of the fluid
to be large enough that radiation losses can be neglected
(e.g. \citealt{Giannios06,Ito13}). The novelty of the code is that of
performing the radiation transfer on the fast evolving background of a
relativistic outflow \citep{Ito15}.  The code injects photons in
thermal equilibrium with the fluid well inside the
photosphere\footnote{For the simulations shown in this work, the
  injection was performed at a Thomson optical depth $\tau_T\sim100$
  or more.} and performs repeated Compton and inverse Compton (IC)
scattering off electrons that are at the local comoving temperature of
the fluid. The present version of MCRaT takes two-dimensional
simulations as input, but the photon propagation is done in three
dimensions to properly take into account the photon propagation angle
with respect to the fluid. In other words the photon field includes
photons propagating in directions with non-zero azimuthal angle
$\phi$, even if the fluid velocity is constrained to be only in the
radial and polar direction. In more detail, the code performs the
following steps (see also Figure~\ref{fig:mcrat_blocks}):
\begin{itemize}
\item {\bf Photon injection --} Photons are injected at a user
  specified radius $R_\mathrm{inj}$ and within a user specified range
  of polar angles from the jet axis. The code determines the local
  temperature, density, and velocity vector at the photon's position
  by nearest-neighbor interpolation. This can be switched to linear
  interpolation at the cost of a longer computational time, but with
  no visible effect on the results. The photon 4-momentum
  ${\mathbf p_{\gamma}}$ is generated by randomly sampling a thermal
  photon distribution at the local fluid temperature. The distribution
  can be chosen between a black body and a Wien spectrum, the former
  being appropriate for fireballs in the acceleration stage, the
  latter for fireballs in the coasting phase. The photons' direction of
  propagation is assumed to be isotropic in the fluid's rest frame at
  injection. Since it is clearly impossible to simulate all the
  photons in a GRB fireball (their number amounting to $\sim10^{59}$
  for a typical burst), each photon is assigned a weight:
\begin{equation}
  w_{\gamma,i}=\frac{8\pi \, c}{3N_{\rm{inj}}} \xi \, T^{\prime 3}_i \, \Gamma_i \, R_i^2
  \, \delta t \,\delta\theta \,\sin(\theta_i)
\label{eq:weight}
\end{equation}
where $\xi$ is the coefficient for the photon number density
($n_\gamma$) in the equation $n_\gamma=\xi T^3$; $\xi=20.29$ for a
black body and $\xi=8.44$ for a Wien spectrum; $T^\prime$ is the
(comoving) fluid temperature at the photon location; $\Gamma_i$ is the
fluid's bulk Lorentz factor at the photon location; $R$ is the
injection radius (in the Lab frame); $\delta t$ is the time resolution
of the numerical simulation (time interval between two saved frames);
$\delta\theta$ is the polar angular range in which the photons are
injected (in the Lab frame); $\theta_i$ is the specific angle at which
the $i$-th photon is injected (in the Lab frame); and $N_{\rm{inj}}$
is the number of photons injected in each frame of the simulation. The
subscript $i$ is used for all quantities that are computed
independently individually for each photon. The weights are chosen so
that the sum of the energies of the injected photons times their
weight gives the correct jet luminosity (see the Appendix for a
derivation of the photons weight).

\item {\bf Photon mean free path --} The first step in the photons
  propagation is the calculation of their mean free path in the
  expanding fluid. This is computed as \citep{Abramowicz91}:
\begin{equation}
\lambda_i=\frac{dr}{d\tau_T}=\frac{1}{\sigma_T \, n_i^\prime \,
  \Gamma_i \, \left(1-\beta_i \cos\theta_{\mathrm{fl},i}\right)}
\end{equation}
where $\sigma_T$ is the Thomson cross-section, $n_i^\prime$ is the
comoving lepton number density of the fluid at each photon position,
$\beta_i$ is the fluid's velocity at the photon position in units of
the speed of light $c$, and $\theta_{\mathrm{fl},i}$ is the angle
between the fluid and photon's velocities. Note that the mean free
path should in principle be integrated over the evolving fluid density
and velocity. However, photons very rarely move out of their original
hydrodynamic resolution element within the time span $\delta t$
between a scattering and the subsequent one, and therefore our
approximation that the underlying hydrodynamic does not change between
the times at which the mean free path is calculated is accurate. We
also notice that we are assuming that the photon energies are well
below the electron's rest mass energy (511 keV) and we can ignore
Klein-Nishina effects. In most cases, this is a good approximation for
a GRB jet in the comoving frame.

\item {\bf Selection of the interacting photon --} Once all the mean
  free paths have been calculated, a random scattering time for each
  photon is drawn from the distributions:
\begin{equation}
p_i(t)=\frac{c}{\lambda_i}\, e^{-\frac{c}{\lambda_i}t}
\end{equation}
The shortest of the obtained times is used as the next scattering time.

\begin{figure}
\includegraphics[width=\columnwidth]{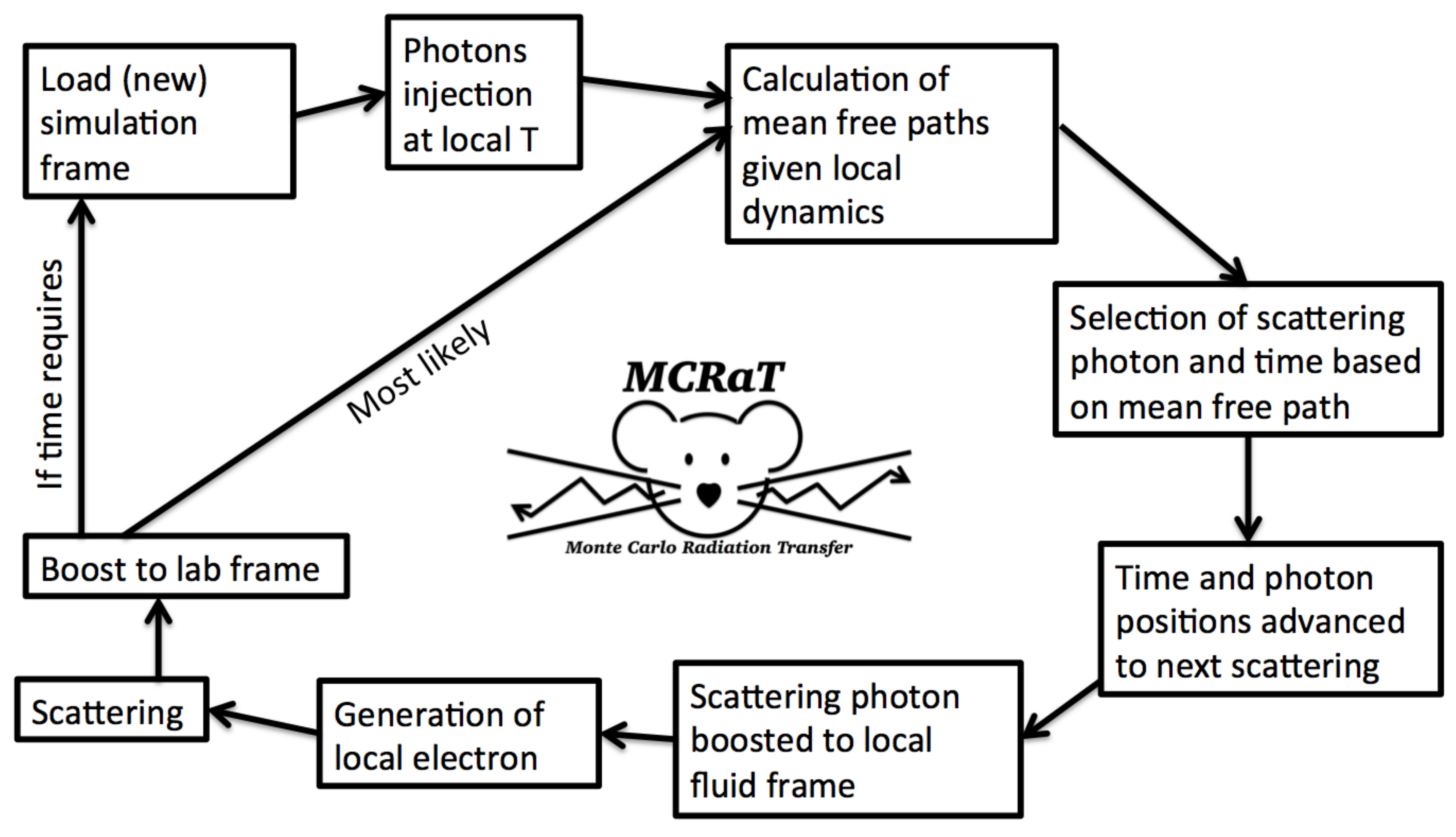}
\caption{{Block diagram of the main loop of the MCRaT code.}
\label{fig:mcrat_blocks}}
\end{figure}

\item {\bf Update of photon positions --} At this point the code
  checks whether the selected collision time is within the time range
  of the current HD simulation frame. If so, all photons positions
  three-vectors $\mathbf{R_i}$ are updated by propagating the photons
  at the speed of light for the selected time interval. If not, a new
  time interval equal to the time remaining for the current HD
  simulation frame is selected, and all photons positions are
  updated. No scattering takes place in this case, a new HD frame is
  loaded, and the code returns to the mean free path step, computing a
  new mean free path for the photons in the updated HD and then
  selecting a new scattering time. Note that, when a new simulation
  frame is loaded, a new batch of photons is added at the injection
  radius following the rules of photon injection as described above in
  bullet ``photon injection''.

\begin{figure}
\includegraphics[width=\columnwidth]{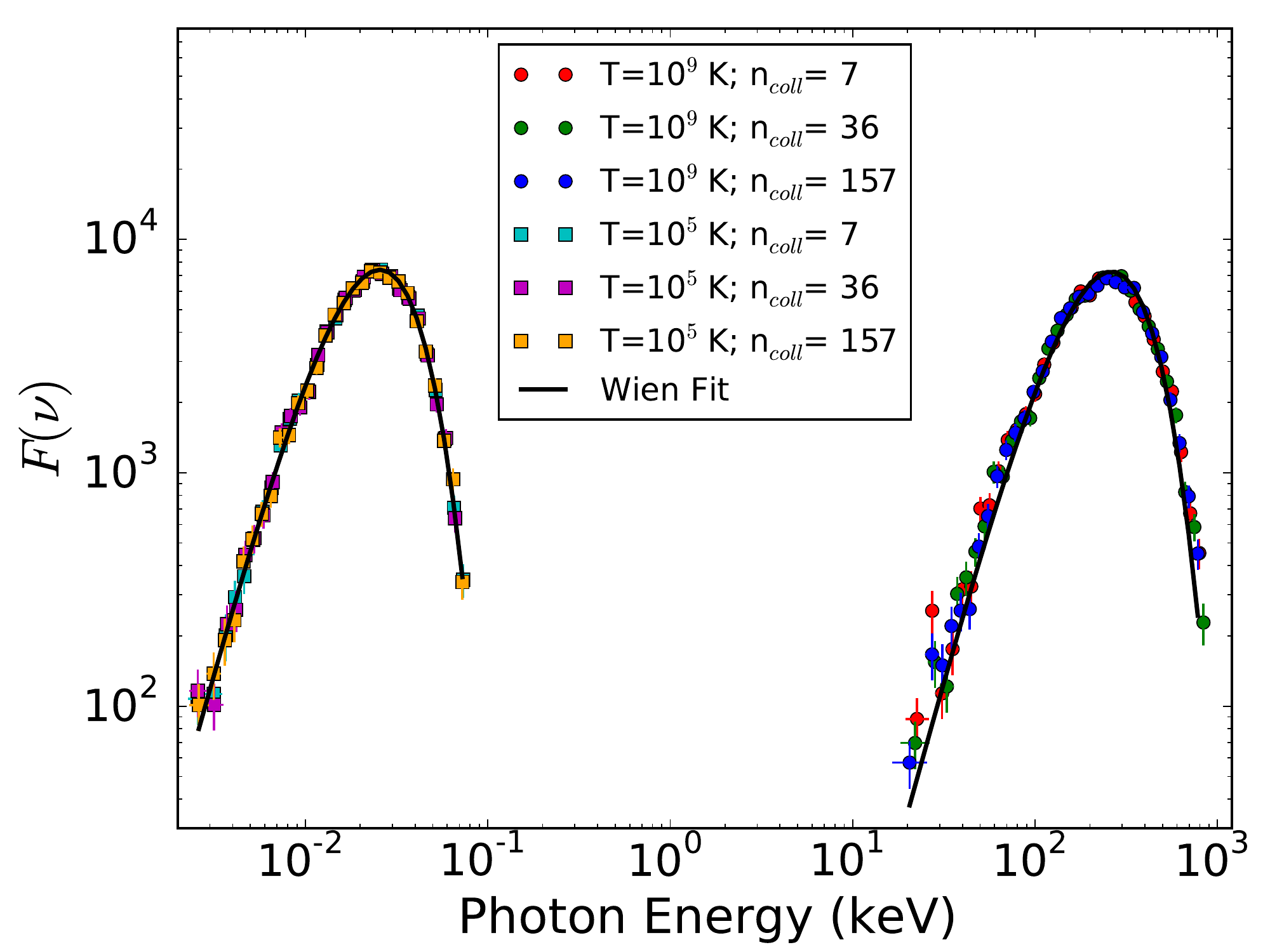}
\caption{{Spectra from a test simulation with a cylindrical outflow of
  constant velocity and temperature. Two simulations are shown, each
  at three different stages, for increasing number of scatterings per photon.}
\label{fig:cyl}}
\end{figure}

\item {\bf Photon scattering --} When the time and location of a
  photon scattering is identified, the code performs several
  steps. First, the photon four momentum $\mathbf{P_i}$ is transformed to
  the local fluid comoving frame. Second, an electron is randomly
  drawn from a Maxwell-Boltzmann or Maxwell-J\"uttner distribution at
  the local fluid temperature $T_i$. The electron is assigned a
  direction of motion with a probability distribution
  $p(\psi)\propto\sin\psi(1-\beta\cos\psi)$, where $\psi$ is the angle
  between the photon and electron velocity vectors.  Third, the
  electron and photon four momenta are transformed to the electron's
  comoving frame, and the reference frame rotated so that the photon's
  propagation is along the z-axis. A scattering angle is generated
  according to the Thomson probability distribution
  $p(\theta_\mathrm{scatt})\propto\sin
  \theta_\mathrm{scatt}(1+\cos^2\theta_\mathrm{scatt})$
  and the post-scattering four momenta are computed according to
  momentum and energy conservation. Finally, the photon's four
  momentum is rotated and doubly boosted back to the lab frame.
\end{itemize}

The loop described above is performed iteratively until the photons
reach the edge of the simulation, or the last simulation frame is
reached (see also Figure~\ref{fig:mcrat_blocks}). Each time a new
simulation frame is loaded, the code saves the photon position vectors
and momenta four-vectors. In post processing, a script is used to
generate a photon event table that contains the photons detected by a
virtual detector placed at a certain distance from the burst and along
a selected orientation. The event table contains photon detection time
and energy and the photon weight $w_{\gamma,i}$.

\begin{figure}
\includegraphics[width=\columnwidth]{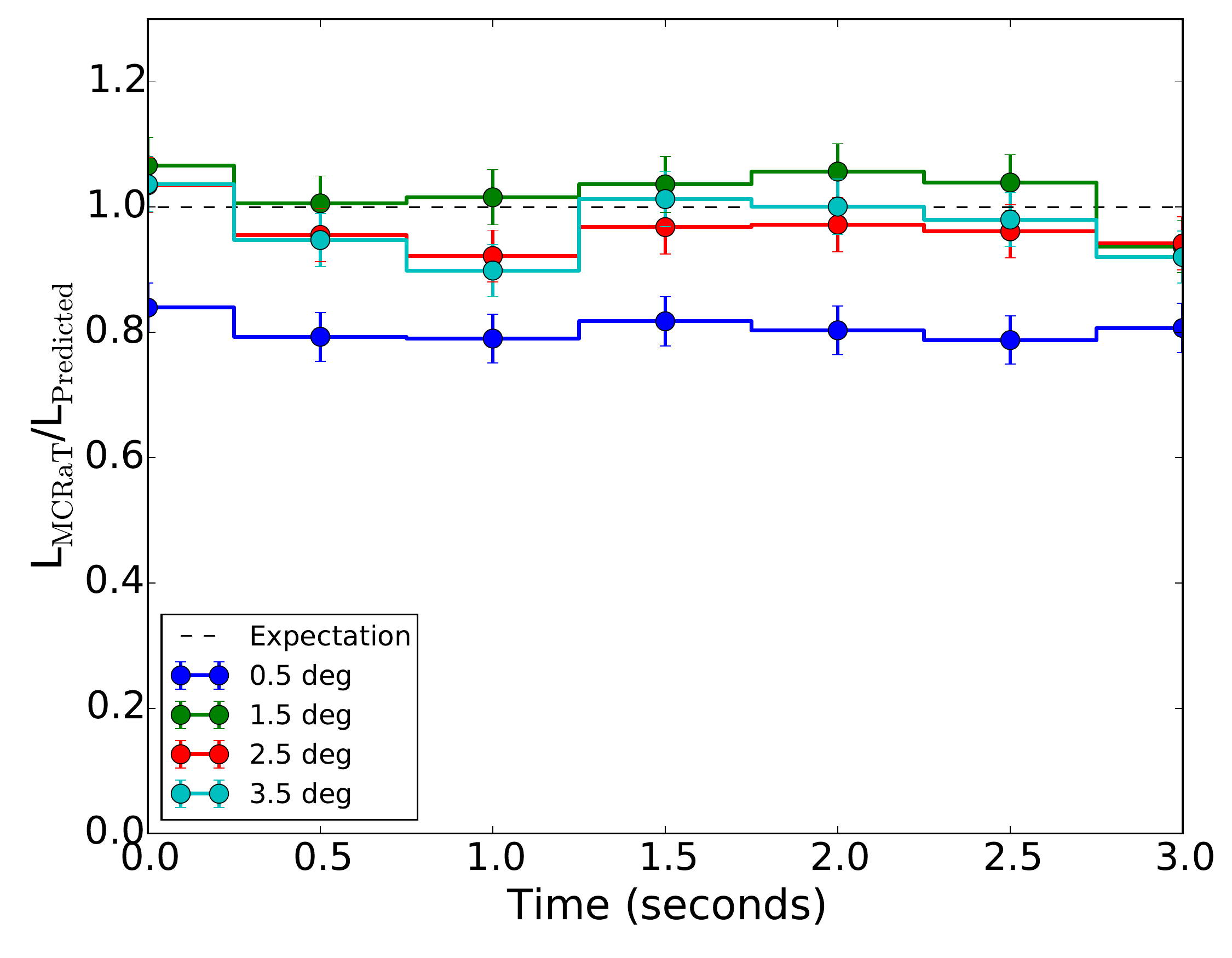}
\caption{{Comparison between the expected luminosity and the one
    computed with MCRaT for a spherical relativistic outflow. For
    the line of sight very close to the singular polar axis we find a
    $\sim20$ per cent deviation between the expected and obtained
    results. All the other lines of sight are fully consistent with
    the predicted luminosity.}
\label{fig:lcsph}}
\end{figure}
 
\subsection{Code Validation}
\label{sec:validation}

In order to validate the MCRaT code we ran a suite of tests for which
the expected result is known. All tests were run by imposing an
analytic outflow solution on the same simulation frames that will be
used in Section~\ref{sec:results} to study the light curves and
spectra of a long GRB simulation. This means that the grid setup (cell
centers and sizes) was read from the simulation files, but the values
of density, pressure, and velocity were substituted with an analytical
solution as a function of the cell position.  This way, the same
spatial resolution of the tests is used for the scientific simulation
and we can evaluate the influence of the grid size and resolution on
the accuracy of the results.

\begin{figure}
\includegraphics[width=\columnwidth]{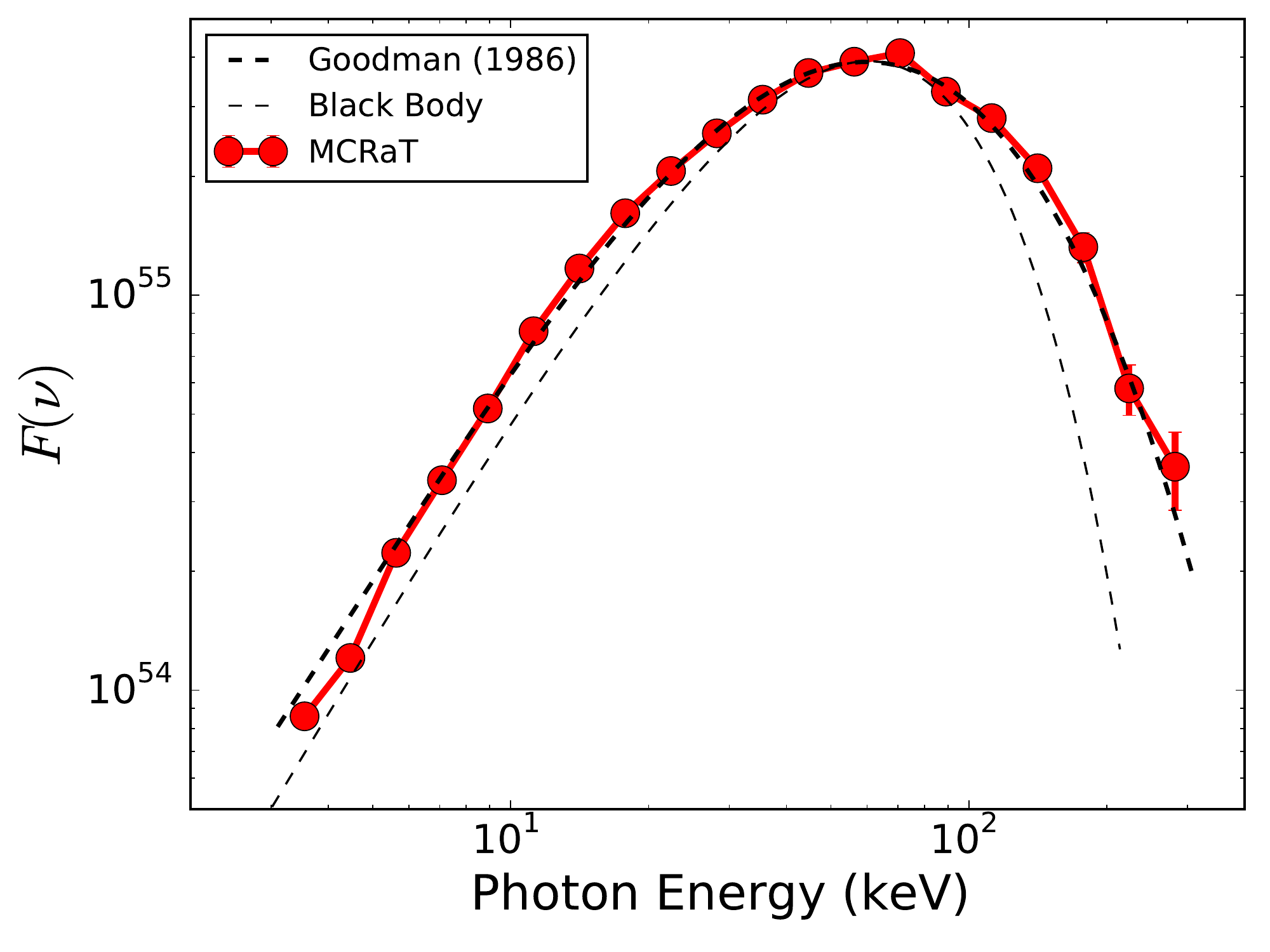}
\caption{{MCRaT spectrum from a spherical outflow compared with a black
    body spectrum at the same temperature and with the prediction of
    \cite{Goodman86} for a spherical relativistic fireball.}
\label{fig:sph_spex}}
\end{figure}

First we run a test in which the outflow is cylindrical. The velocity
is constant and directed along the y axis, the temperature and the
density are constant. We run several of these simulations with
different outflow velocity and/or different temperatures.  Photons
were injected with an initial Wien spectrum. Figure~\ref{fig:cyl}
shows the spectra for two outflow temperatures at various stages
during the simulation, i.e., after increasing number of
scatterings. As expected, the resulting spectra are very well fit by a
Wien spectrum (e.g. \citealt{Chhotray15} and references therein), both
at low temperature (leftmost spectra, non-relativistic electrons) and
at high temperature (rightmost spectra, trans-relativistic
electrons). The figure shows the result from a static configuration
with $v=0$. Analogous results were obtained for outflows with
relativistic outward velocities ($\Gamma=10$, 100, and 300 were
tested).
 
\begin{figure}
\includegraphics[width=\columnwidth]{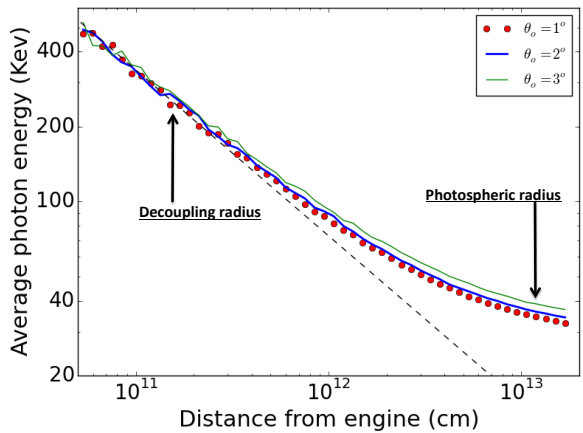}
\caption{{Radial profile of the average photon energy from a test
    MCRaT simulation with a spherical flow. Red marks and colored
    solid lines show the radial evolution of the photon mean energy at
    the observing angles $\theta_o=1$, 2, and 3 degrees. A dashed
    line shows the average energy of photons in equilibrium with the
    plasma. The location of the photosphere and of the radius at which
    the radiation and plasma decouple according to Giannios (2012) are
    shown with vertical arrows.}
\label{fig:sph}}
\end{figure}

As a second test we run a suite of simulations of spherical outflows,
i.e., outflows with only outward radial velocity component. In this
case, the Lorentz factor increases proportionally to the radius until
all the internal energy is converted into bulk motion, saturation is
reached, and the fireball transitions from accelerating to coasting at
constant Lorentz factor.  In the acceleration stage the electrons'
temperature decreases proportionally to $r^{-1}$; as the asymptotic
Lorentz factor is reached, the fireball coasts at constant $\Gamma$
and the electrons' temperature decreases as $r^{-2/3}$. We stress that
this analytical solution was imposed, we did not run a hydrodynamic
simulation of a spherical outflow, we simply replaced the output of
the simulation with the analytical prescription.

We first compare the photon luminosity of the spherical fireball
calculated from the simulations with the analytical expectation. The
result is shown in Figure~\ref{fig:lcsph}, where the MCRaT luminosity
for four different observers is compared to the analytical
expectation. While for observers away from the singular polar axis we
find good agreement, a deviation of $\sim20$ per cent is observed at
$\theta_0=0.5^\circ$. We ascribe this to the artifact of the polar
axis. Since the polar axis is singular in the coordinate system, the
weight of photons close to the axis can be very small
(cfr. Eq.~\ref{eq:weight}). For infinite number of photons, all small
angles are sampled and there is no systematic deviation in the
results. For a finite number of photons, however, significant
non-Poissonian deviations can result from under- or oversampling the
smallest polar angles. An analogous effect can be seen in Figure 8 of
Morsony et al. 2007 (dashed line) where an unphysical drop of the
luminosity is observed at small polar angles from the jet axis. In the
remainder of this manuscript we will compute light curves and spectra
only for photons propagating with a polar angle of at least
$0.5^\circ$ to avoid this problematic region.

In Figure~\ref{fig:sph_spex} we show the time-integrated spectrum
accumulated between angles $\theta_o=1$ and 2 degrees from the polar
axis\footnote{Note that since each photon propagates along a unique
  direction, light curves and spectra cannot be calculated at a
  precise viewing angle. Rather, photons within a small but non-zero
  range of angles need to be added. In the manuscript, we accumulate
  spectra and light curves from photons within an acceptance of 1
  degree, and use the average of these angles to indicate the
  direction of observation.}. The spectrum is broader than a Planck
function, and agrees very well with the prediction of
\cite{Goodman86}. Finally, we show in Figure~\ref{fig:sph} the radial
dependence of the average photon energy for various off-axis angles
compared to the average frequency of photons in equilibrium with the
fluid.  As described in great detail in Beloborodov (2011; see
e.g. their Figure~3), the photon temperature follows the electron
temperature at large optical depth $\tau>10$. As the photosphere is
approached, the electrons and photons decouple and the photons start
to cool at a slower rate until, at the photosphere, the decoupling of
the radiation is complete and the photon temperature remains
constant. The MCRaT result is shown in Figure~\ref{fig:sph}. Red
symbols and colored lines show the average photon energy for viewing
angles $\theta_o=1$, 2, and 3$^\circ$. A dashed line shows instead the
average energy of photons strongly coupled to the electrons. Two
vertical arrows show the location of the photosphere and the location
at which the \cite{Giannios12} condition is attained (the Wien radius,
\citealt{Beloborodov13}). The MCRaT result reproduces all the features
of the analytical solution \citep{Beloborodov11} and confirms that the
approximations adopted so far for computing the peak energy of the
prompt GRB spectrum from simulations are inaccurate. While it is true
that electrons and photons start decoupling at or around the Wien
radius, the decoupling is incomplete and the spectrum keeps cooling by
interaction with the electrons.  This approximation was adopted in L13
and as a consequence their results overestimate both the light curve
luminosity and the peak frequency of the spectrum. On the other hand,
the assumption that the electrons and photons are coupled until the
photosphere is reached is also incorrect. This approximation, adopted
by \citep{Lazzati09,Lazzati11,Mizuta11,Nagakura11} underestimates the
radiation peak frequency and the light curve luminosity. A correct
implementation of the radiation transfer in numerical simulations of
the photospheric model is therefore necessary not only to obtain the
spectral properties beyond the peak frequency (the low- and high-
frequency spectral indices), but also to correctly estimate the peak
frequency itself. Before showing the MCRaT results for a GRB
simulation, we stress that within the viewing angles presented in this
paper the code is quite accurate, the photon average energies
differing only by a few per cent for different lines of sight
(Figure~\ref{fig:sph}).

\section{Results}
\label{sec:results}

\begin{figure}
\includegraphics[width=\columnwidth]{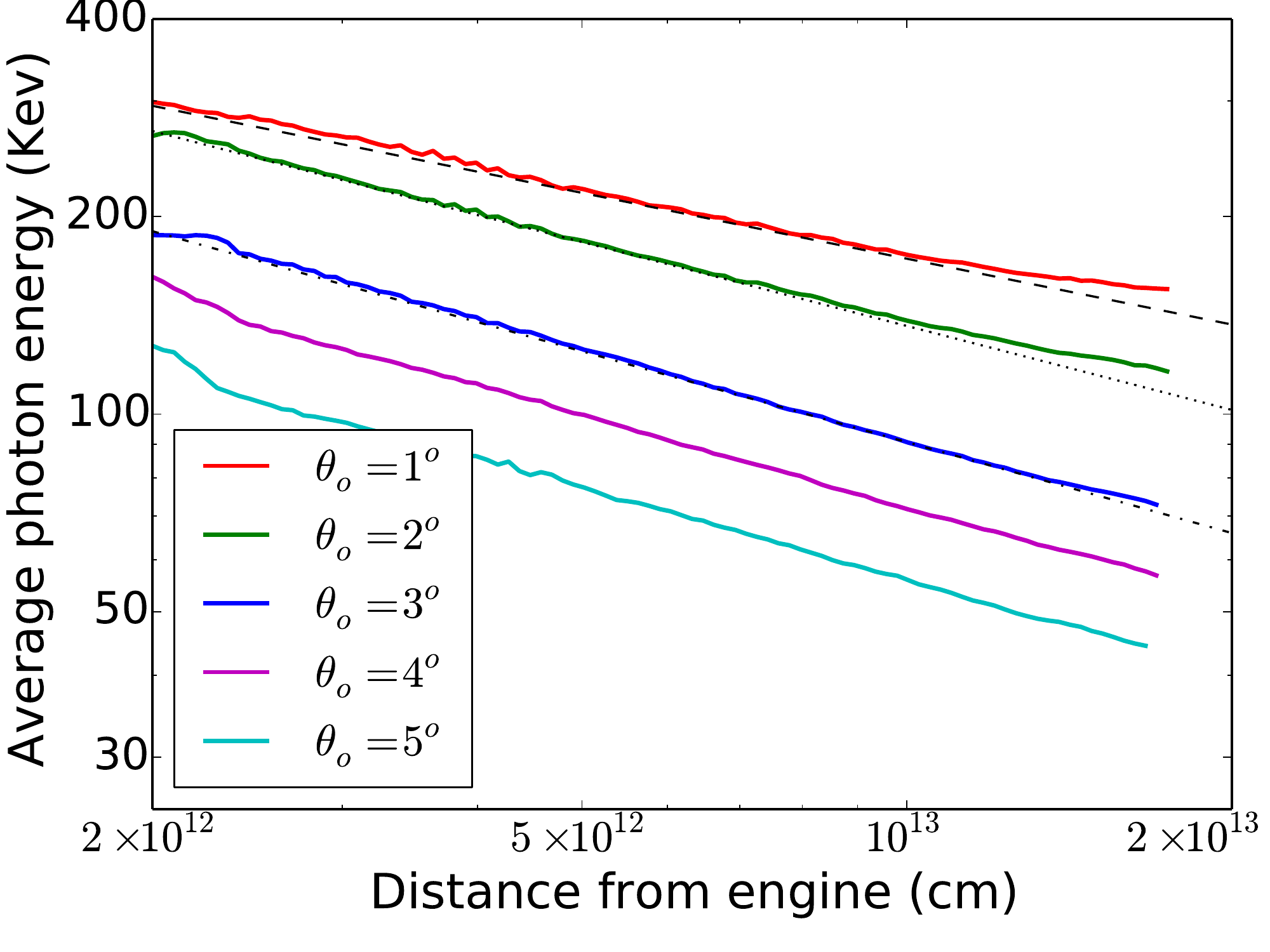}
\caption{{Radial profile of the average photon density for radiation
    propagating at different off-axis angles.  Analogous to
    Figure~\ref{fig:sph} but for the GRB jet simulation. Guidelines
    are plotted for off-axes 1, 2, and 3 degrees, to show the
    deviation from the self-similar behaviour as the photosphere is
    approached (or lack thereof).}
\label{fig:radial16TI}}
\end{figure}

We have applied the MCRaT code to our fiducial simulation of a long
duration gamma-ray burst jet that was previously studied in the
approximation of a sharp photosphere (\citealt{Lazzati09,Lazzati11};
L13). The simulation runs for $\sim600$~s, at which time the jet head
has reached a distrance of $\sim1.3\times10^{13}$ cm from where it was
launched, the center of the progenitor star. The progenitor is modeled
as a 16 solar mass Wolf Rayet star, model 16TI from
\cite{Woosley06}. The jet is injected as a boundary condition at
$r_0=10^9$~cm, with luminosity $L=5.33\times10^{50}$~erg~s$^{-1}$,
half-opening angle $\theta_0=10^\circ$, Lorentz factor $\Gamma_0=5$,
and internal energy to allow for an asymptotic jet acceleration to a
terminal Lorentz factor $\Gamma_\infty=400$. After $t=100$~s, the
engine luminosity is decreased by a factor 1000 to simulate the turn
off of the engine. The radial resolution at $R=10^{13}$~cm, where the
photosphere is expected to lay, is $\delta y=10^{9}$~cm, giving
$\delta y/y=10^{-4}$, comparable to $1/\Gamma^2$. Resolution in the
orthogonal $x$ coordinate is identical since we use square resolution
elements. Shocks at the photospheric radius are therefore poorly
resolved. Still, this resolution is about an order of magnitude better
than in the simulations of \cite{Ito15}. As we discuss below, limited
resolution is a serious concern for these results and an increase by
at least an order of magnitude is required to ensure that shocks are
resolved and that the simulations capture the temperature variations
that they produce. 300,000 photons were injected in the simulation at
off-axis angles $\theta_i\le6^\circ$.

\begin{figure}[!t]
\includegraphics[width=\columnwidth]{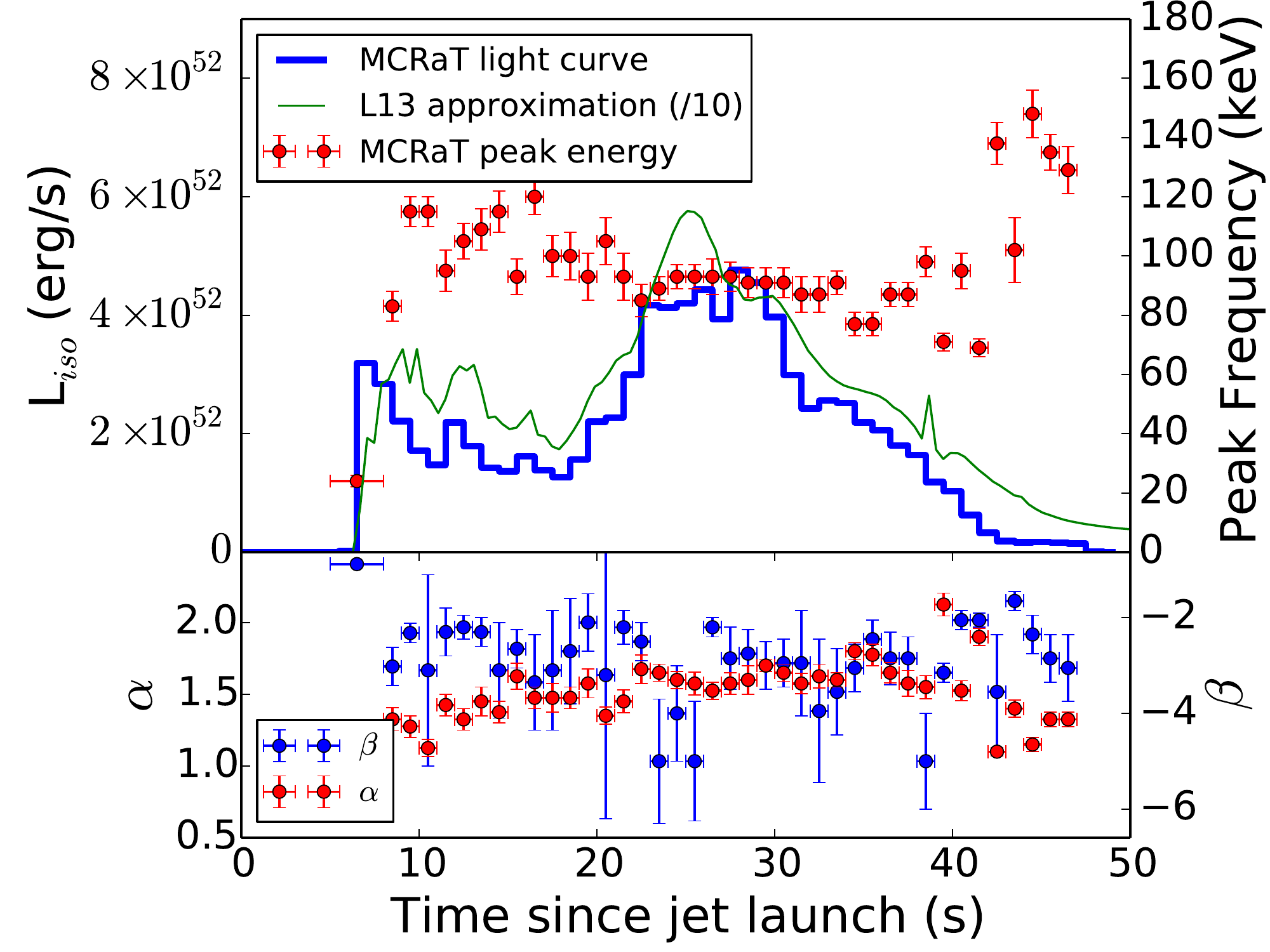}
\caption{{Light curve and spectral evolution for photons traveling
    within the acceptance range $0.5\le\theta_{\rm{ph}}\le1.5$ (for an
    observer laying at $\theta_o=1^\circ$) from the polar axis. In the
    upper panel, the thick solid line shows the MCRaT result, the thin
    line shows the result from L13 (rescaled by a factor 10) and the
    symbols show the peak energy in 1 second temporal bins (energy
    y-axis to the right). In the bottom panel, the Band parameters
    $\alpha$ and $\beta$ are shown with red and blue symbols,
    respectively. The $\alpha$ scale is to the left, while the $\beta$
  scale is to the right.}
\label{fig:theta1}}
\end{figure}

Under conical expansion, the jet we input in the simulations would
become optically thin (and therefore have its photosphere) at
$R=6.4\times10^{11}$~cm (e.g., \citealt{Meszaros00}). However, the
progenitor-jet interaction induces a non-conical expansion and
prevents the saturation of the acceleration at $\Gamma=400$, resulting
in a much larger photospheric radius. \cite{Lazzati11} found a
photospheric radius of the order of $R=10^{13}$~cm in the same
simulation, increasing for larger off-axis angles. With MCRaT we do
not compute the photospheric radius by integrating the optical depth
(as is done in all previous numerical simulations of photospheric
emission in GRBS); we follow instead the photon propagation. Whether
the photons have crossed the photospheric radius or not can be
understood by looking at the radial dependence of their average
energy. The average photon energy decreases at radii smaller than the
photosphere and attains an asymptotic value beyond it
(\citealt{Beloborodov11}; see also Section~\ref{sec:validation}). In
Figure~\ref{fig:radial16TI} we show the time-averaged MCRaT results
for off-axis angles of 1, 2, 3, 4, and 5 degrees, each with an
acceptance of $\pm0.5$ degrees. Analogously to Figure~\ref{fig:sph},
dashed lines are overlaid to show the average frequency of photons
coupled to the electrons. The colored lines detach from the dashed
ones at the Wien radius and become constant beyond the photospheric
radius. In agreement with results obtained from integration over the
optical depth (\citealt{Lazzati11}, see their Figure 2), we find that
the photosphere lies at $R_{\rm{ph}}\sim1.5\times10^{13}$~cm for small
off-axis angles but is located beyond the outer edge of the simulation
box (which is located at $2.5\times10^{13}$~cm) for relatively large
off axis angles $\theta_o>3^\circ$. In mode detail, the photosphere is
clearly reached for $\theta_o=1^\circ$, is almost reached for
$\theta_o=2^\circ$, and it is approached for $\theta_o=3^\circ$. In
the following we show results for these three off-axis angles, keeping
in mind that the result for $\theta_o=3^\circ$ might be approximate
since the simulations do not appear to be extended enough to
completely reach the photosphere. It should be remembered, however,
that the photosphere is not at a constant distance throughout the
simulation, starting at a large radius and progressively closing in
(see, e.g., the blue lines in Figure 4 of L13). Even for the larger
off-axis angles, the photosphere is likely reached at some time after
the initial high-density plug at the jet's head is overcome.

\subsection{Light Curves and Spectra}

\begin{figure}[!t]
\includegraphics[width=\columnwidth]{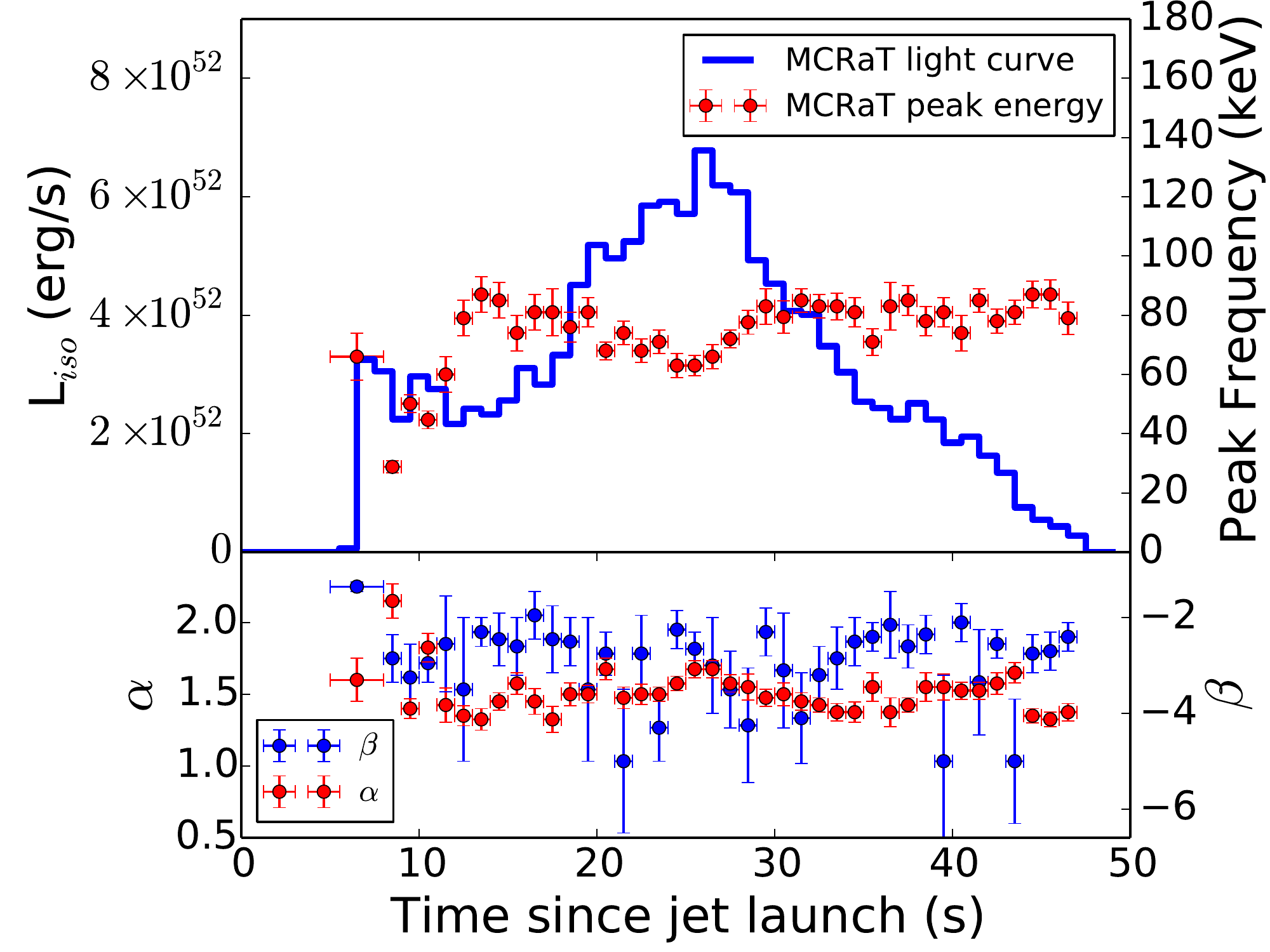}
\caption{{Same as Figure~\ref{fig:theta1} but for an acceptance angle
    $1.5\le\theta_{\rm{ph}}\le2.5$. L13 results at this angle are not
    available for comparison.}
\label{fig:theta2}}
\end{figure}

\begin{figure}[!t]
\includegraphics[width=\columnwidth]{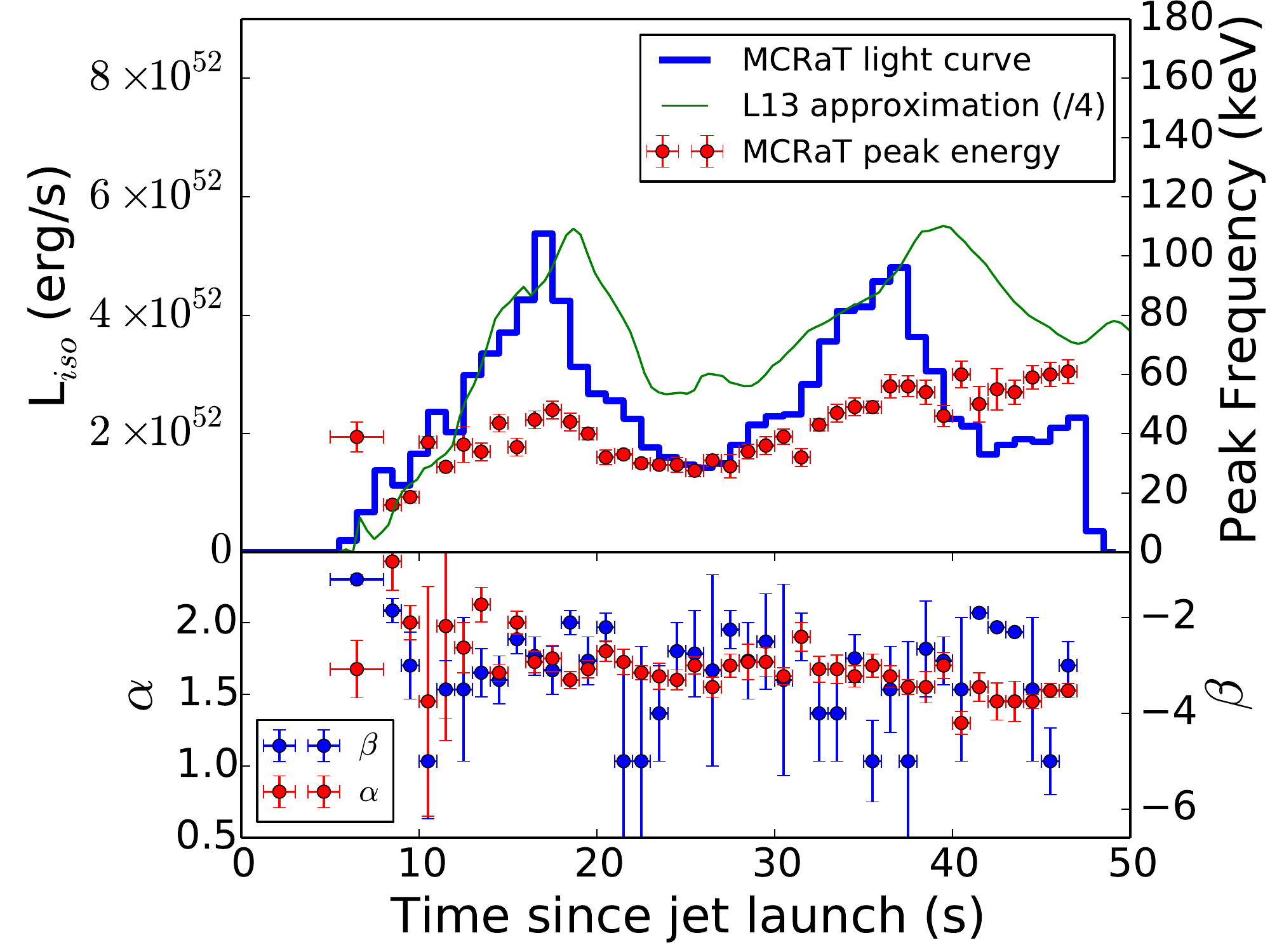}
\caption{{Same as Figure~\ref{fig:theta1} but for an acceptance angle
    $2.5\le\theta_{\rm{ph}}\le3.5$.}
\label{fig:theta3}}
\end{figure}

Light curves and spectra are computed by binning the photons within
the acceptance angle in time and/or frequency, each multiplied times
its weight (Eq.~\ref{eq:weight}). Light curves for
$0.5\le\theta_{\rm{ph}}\le1.5$, $1.5\le\theta_{\rm{ph}}\le2.5$, and
$2.5\le\theta_{\rm{ph}}\le3.5$ degrees are shown as thick lines in
Figures~\ref{fig:theta1}, \ref{fig:theta2}, and~\ref{fig:theta3},
respectively. The bin time is one second in all cases. In
Figures~\ref{fig:theta1} and~\ref{fig:theta3} a thin line shows the
light curves computed with the sharp last energy surface approximation
by L13 (see also \citealt{Giannios12}). Normalization factors of 10
and 4 were applied to the two curves, respectively, to facilitate the
comparison. The MCRaT curves have similar temporal evolution compared
to the old sharp approximation. This is not surprising since the light
curve shape is mainly due to the hydrodynamic effect of recollimation
that the progenitor star imprints on the jet.  Such effect is purely
hydrodynamic in nature and is therefore independent of how the photon
propagation is treated in post-processing. The difference in
normalization is due to the fact that, as seen in
Figures~\ref{fig:sph} and~\ref{fig:radial16TI}, the average photon
frequency decreases after the last energy exchange radius assumed in
L13.

\begin{figure*}[!t]
\includegraphics[width=\textwidth]{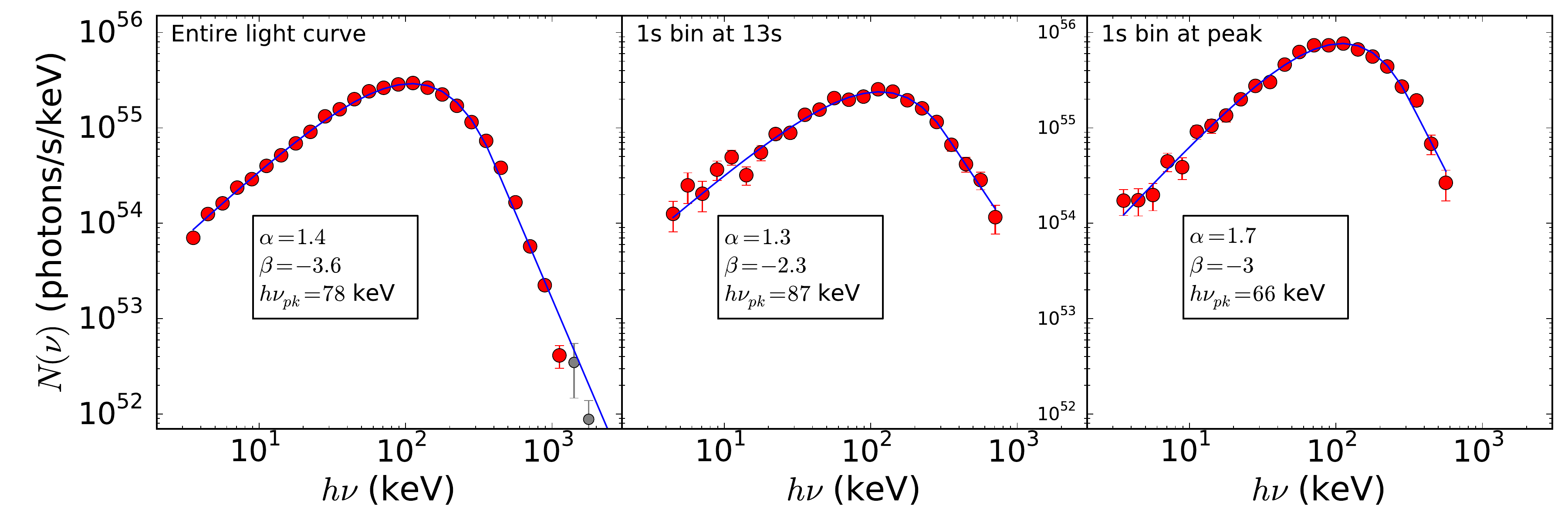}
\caption{{Sample MCRaT photon-count spectra for the
    $1.5\le\theta_{\rm{ph}}\le2.5$ observer. The left panel shows the
    spectrum accumulated over the entire light curve and fit with a
    Band model. The central panel shows one of the most non-thermal
    spectrum, accumulated over a 1 s time bin at $t=13$~s. Finally,
    the left panel shows the spectrum at the light curve peak. Each
    panel shows the best fit Band parameters in an inset and the best
    fit Band spectrum as a thin solid line. In the left panel bins
    with some photon counts but not enough statistics to be included
    in te fit are shown in gray and with a smaller size.}
\label{fig:spex}}
\end{figure*}

A few example spectra for the $1.5\le\theta_{\rm{ph}}\le2.5$ case are
shown in Figure~\ref{fig:spex}. We have chosen this case because it is
far enough from the jet axis to avoid spurious effects from the
reflective symmetry along the y-axis. In addition, as shown in
Figure~\ref{fig:radial16TI}, the photospheric radius is contained in
our simulations at this off-axis angle. The three panels show the
photon count spectra for the whole burst (left panel), a fairly
non-thermal time interval (central panel), and the most luminous time
bin of the light curve (left panel). In all cases we notice some
features in agreement with observed gamma-ray bursts and others in
significant tension with the observations. All spectra are fit with a
Band model \citep{Band93} and the best values for the parameters
$\alpha$, $\beta$, and $h\nu_{\rm{pk}}$ are reported.  In the Band
model, a smoothly joined broken power-law function, $\alpha$ is the
asymptotic photon index at low frequency, $\beta$ is the asymptotic
photon index at high frequency, and $h\nu_{\rm{pk}}$ is the photon
energy at which the $\nu F(\nu)$ spectrum peaks.

Going from low- to high-frequencies, we first notice that the
low-energy spectral index $\alpha$ is found to be significantly larger
than in observations. Average observed values of $\alpha$ range
between -1 and -0.5 \citep{Preece00,Goldstein13,Gruber14,Yu16} while
we consistently find $\alpha>1$. This is not surprising since the
difficulty of the photospheric model to reproduce the observations in
the low-frequency range is well known. Our synthetic spectra have
power-law indices even steeper than a black body, due to the fact that
the spectrum was injected with a Wien distribution of
frequencies\footnote{The low-frequency photons require more scattering
  to reach equilibrium and our vary soft tail may also be affected by
  an insufficient number of scattrings \citep{Chhotray15}.}. Injecting
a black body spectrum would have made the low-frequency spectrum
harder, however the steepening of the spectrum would not have bee
substantial enough to reproduce the typical spectra of GRB
observations. We also notice that our simulations do not test any of
the solutions proposed to solve the discrepancy. For example, models
based on sub photospheric dissipation (e.g. \citealt{Chhotray15}) can
be tested only with simulations of higher resolution than the one
presented here, since shocks need to be fully resolved to affect the
radiation spectrum. Models based on the addition of soft photons from
synchrotron \citep{Vurm11,Vurm16} require instead a radiation transfer
code that allows for photon emission and absorption, beyond the
current capabilities of MCRaT.

The peak energy of the light curves is found in the $\sim100$~keV
range. This is a fairly common value for a burst of intermediate
luminosity like the one simulated here. Yet, it is on the low tail of
the distribution and we will see in Section~\ref{sec:correls} that
there is some strain between our results and the Amati and Yonetoku
correlations.

The best agreement between our results and observations is found at
high frequencies. We first notice that we can exclude both a cutoff
power-law and a Black Body fit withs high confidence. A $\chi^2$-based
F-test yields probabilities $P<10^{-5}$ for the two simpler functions
mentioned above to give a comparable fit with respect to a Band
Function. A high-frequency power-law is therefore required by the
data, as in observed bursts. In addition, we find that MCRaT's $\beta$
values, between 2 and 3, are fairly typical for long-duration GRBs
\citep{Preece00,Goldstein13,Gruber14,Yu16}. We note that the spectra
shown only extend to $\sim1 MeV$ due to poor statistics, but there is
no indication of a cutoff. In the left panel -- the one with the best
statistics -- we show with smaller symbols and in grey color the data
in the spectral bins that have too little photons to be included in
the fit. We notice that these data points are consistent with the
extrapolation of the high-frequency power-law and no break or cutoff
is required. Very high frequency photons in the GeV band detected by
Fermi \citep{Abdo09} are however better explained as forward shock
emission in the photospheric model \citep{Kumar09,Ghisellini10}.

Figures~\ref{fig:theta1}, \ref{fig:theta2}, and~\ref{fig:theta3} show,
besides the light curves, the results of time-resolved
spectroscopy. Photons were accumulated in 1 s bins and a Band spectrum
was fit to each temporal bin and for each of the three off-axis
observers. The top panel of each figure reports the peak energy in keV
(right hand vertical scale), while the bottom panel reports the best
fit $\alpha$ and $\beta$ values. The low-frequency index $\alpha$
refers to the left hand vertical scale, while the high-frequency index
$\beta$ refers to the right hand vertical scale. We first notice that
there is quite some diversity in the kind of spectral evolution
observed in the three figures. At small off-axis $\theta_o=1^\circ$,
the peak frequency follows a general hard-to-soft trend, while the
low-frequency index $\alpha$ has a small growing trend.  Eventually,
both trends are broken. At times larger than $t\sim40$~s the peak
energy rises above 100~keV while the spectral index $\alpha$
fluctuates with no stable trend. This transition is due to the
emergence of the unshocked jet, as already seen in previous
simulations (\citealt{Morsony07,Lazzati09}; L13). The hard-to-soft
trend has been observed in some bursts. The $\alpha$ increase, on the
other hand is not observed but seem irrelevant since, as already noted
above, our spectra have $\alpha$ indices in significant disagreement
with observations.

Moving away form the axis, at $\theta_o=2^\circ$ we find a light curve
with very little spectral evolution, except for an initial hardening
and a small softening around the peak. This is quite unusual in
observations. Finally, at $\theta_o=3^\circ$ we find an example of
tracking behavior, the peak energy increasing at high luminosity and
decreasing at low luminosity. Again, towards the end of the burst a
global hardening is observed, possibly due to the emergence of the
unshocked jet or to the fact that photons at the end of the burst
haven't yet reached the photosphere. In all three cases, non-thermal
high-energy tails are measured throughout the burst, with the
exception of a few intervals in which the high-frequency index was
fixed to $\beta=-5$. 

We stress that the jet input in the simulation has constant luminosity
and any variation in the light curve luminosity is brought about by
the jet-progenitor interaction. There is evidence that variability is
injected at the base of the jet \citep{Morsony10}. More pronounced
spectral evolution is expected for burst with variable central engines
\citep{Lopez14}. In addition, the simulations we present here might
have engines that are active for too long a time, as pointed out in
\cite{Lazzati13b}.

\subsection{Correlations}
\label{sec:correls}

\begin{figure}
\includegraphics[width=\columnwidth]{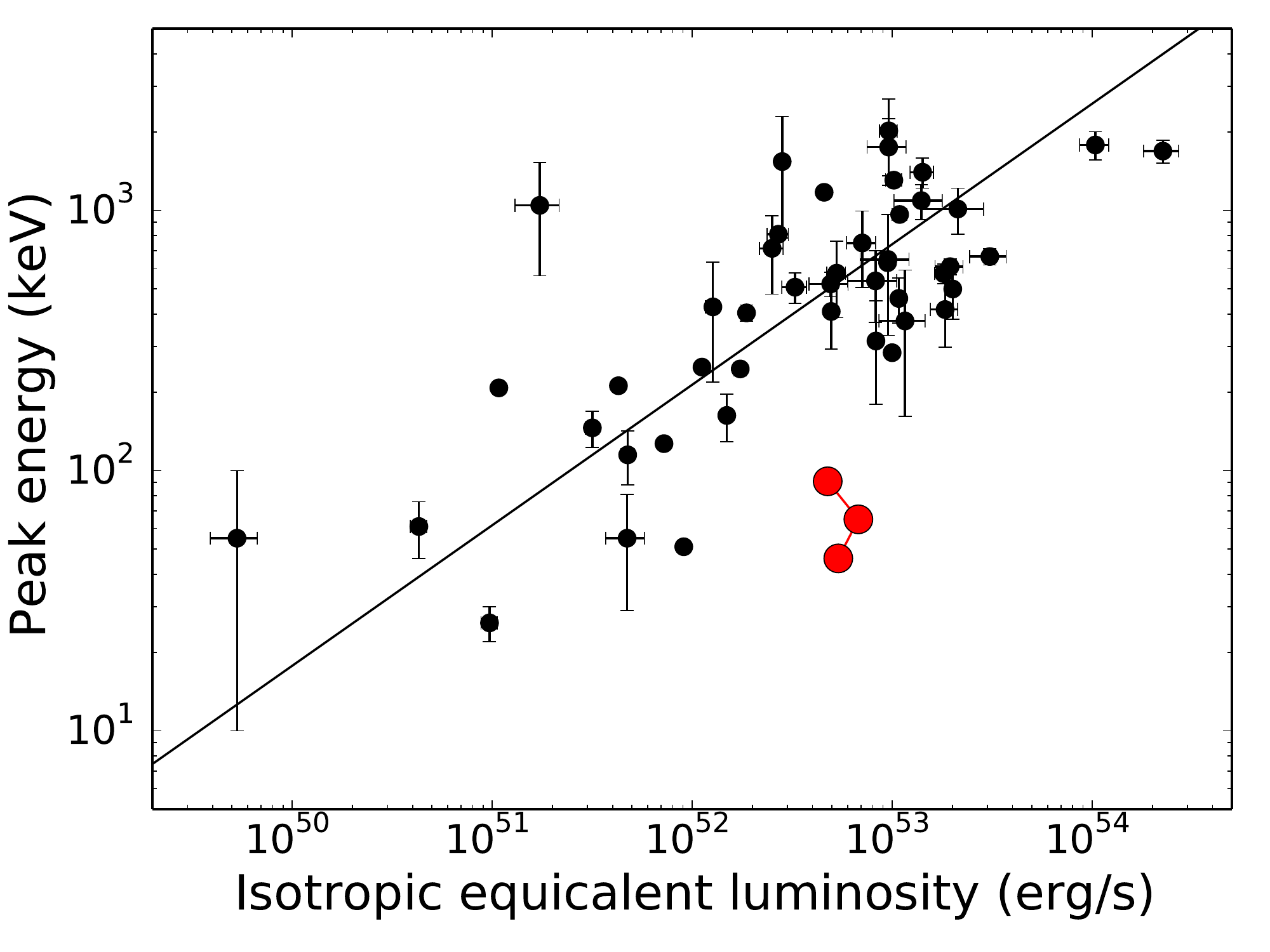}
\caption{{Comparison between the observational Yonetoku correlation
    (black symbols with error bars and best fit solid line) with the
    MCRaT results (red symbols). }
\label{fig:yone}}
\end{figure}

One of the reasons for the success of the photospheric model was its
ability to reproduce the Amati, Yonetoku, Golenetskii,
$E_{\rm{iso}}-\Gamma$, and energy-efficiency correlations
\citep{Golenetskii83,Amati02,Yonetoku04,Liang10,Ghirlanda12}. This was
demonstrated with the results of numerical simulations, and the
scaling was ascribed principally to a line of sight effect (L13). With
MCRaT we can re-analyze the success of the photospheric model in
reproducing the observational ensemble correlations. We show in
Figure~\ref{fig:yone} the result for the Yonetoku correlation between
the 1-s binned peak luminosity and the peak frequency measured at the
time of the peak (again, integrated over 1~s). Given the results shown
above, it does not come as a surprise that the MCRaT result are not a
perfect match to the observations. The assumption of a complete
decoupling at the Wien radius adopted by L13 (see also
\citealt{Giannios12}) is indeed fairly crude. While it is true that
the radiation decouples from the leptons at that optical depth, the
radiation continues to cool, even though remaining at a temperature
that is higher than that of the matter (\citealt{Beloborodov11}; see
also Figure~\ref{fig:sph}). As a consequence, the MCRaT spectra peak
at a frequency that is too low by a factor $\sim3$ with respect to
observations. Such an offset is not large enough to reject the model,
since the dispersion of the data is comparable, yet it puts the
simulation results at some strain with the data. Analogous results
hold for all the aforementioned correlations, with the exception of
the $E_{\rm{iso}}-\Gamma$ and energy-efficiency correlations, for
which no significant discrepancy is found. The agreement, however, is
most likely due to the poor definition of the correlations in the data
rather than to a true success of the model.

\section{Summary and discussion}
\label{sec:discussion}

We have presented a novel numerical code to perform Monte Carlo
radiation transfer in relativistic jets under the assumption that the
radiation-matter interaction is dominated by Compton scattering off
the leptons in the fluid. Our code is analogous to the one recently
developed by Ito et al. (2013, 2015)\footnote{The difference between
  the two codes is technical (language, dimensionality) but the
  physics is the same.} and we find similar results, even though they
applied it to a different set of simulations. Neither MCRaT nor the
Ito et al. (2013, 2015) code cannot take into account the radiation
feedback on the hydrodynamics. However, the hydrodynamic simulations
to which the Monte Carlo codes are applied assume infinite coupling
(the equation of state for both photons and plasma are the same) and
therefore the dynamics from the HD simulations should be accurate in
the optically thick part of the outflow, where most of the Compton
scatterings take place and the spectrum takes its shape. Despite these
limitations, both MCRaT and the Ito et al. code constitute a
significant advance over previous studies, in which the coupling
between matter and radiation was assumed to be complete at optical
depths larger than a critical value and null at smaller optical
depths. Said critical optical depth was assumed to be either unity
\citep{Lazzati09,Lazzati11,Mizuta11,Nagakura11}, or of several tens
(\citealt{Giannios12}; L13). As analytically predicted for spherical
outflows by \cite{Beloborodov11}, proper radiation transfer shows that
the radiation and matter indeed start to decouple at an opacity of
several tens. However, the radiation keeps cooling significantly until
the photosphere is reached, at which point the radiation propagates
with constant average properties\footnote{It is envisageable that in
  the presence of a significant dissipation event, such as an external
  shock, the radiation properties can be changed well after the
  photosphere is crossed. Such events are, however, not present in our
  simulations.}. In addition, MC radiation transfer allows for a full
calculation of the emitted spectrum, while only its peak frequency
could be estimated under the approximations previously employed.

We have applied MCRaT to our fiducial GRB jet simulation, used in
previous photospheric work. We find that the MCRaT results predict
light curves that are dimmer and characterized by lower frequency
photons with respect to the results from L13. Spectral and temporal
analysis was performed on the resulting light curves and spectra. We
find that, as predicted by many analytical and semi-analytical
previous publications, photospheric radiation is characterized by a
significant non-thermal high-frequency tail. Fitting the time-resolved
spectra with a Band function we find values of the index $\beta$ in
good agreement with the observations and a significant evolution of
the slope over the burst light curve. Because of the continued
radiation cooling in the sub-photospheric region, we find that the
MCRaT spectra are characterized by photons of lower frequency with
respect to the L13 results. This deteriorates the agreement of the
simulation results with the Amati and Yonetoku ensemble
correlations. While the MCRaT results are still within the observed
dispersion, they do not follow the correlation closely and more work
is necessary to better understand the role of photospheric emission in
the establishment of such correlations. In addition, we find that our
simulations do not reach the photospheric radius for off-axis angles
$\theta_o>2^\circ$, preventing the analysis of the off-axis light
curves that were deemed to be responsible for the correlations
(L13). We also find that the low-frequency spectral index is much
steeper than observed. This is a well known problem of photospheric
emission in the absence of an efficient emission mechanism or of
strong sub-photospheric dissipation and the discrepancy found is not
surprising.

One possible shortcoming of this work that could be at the basis of
the discrepancies discussed above is the resolution of the
simulations. In order to study the effect of shock-heating on the
spectrum we need to be able to resolve the shocks so that their true
temperature can be taken into account. Unfortunately, the resolution
of the simulations presented here is sufficient only to moderately
resolve the shocks. As a consequence, continuous heating from
recombination, or the kind of shock heating envisaged in, e.g.,
Chhotray and Lazzati (2015), could not be studied. We plan to perform
higher-resolution simulations for both long and short GRB jets in the
near future to test the impact of resolution on the results. Some hope
comes from the comparison of our results with those of
\cite{Ito15}. Their resolution is lower than ours, but they consider
precessing jets that substantially increase dissipation. Even though
they do not perform a detailed time-resolved analysis of their
spectra, it is clear from their Figures~3 and~4 that significant
non-thermal features at both ends of the spectrum can be obtained.  In
addition, we stress that the simulation presented here injects a
constant luminosity from the inner engine and does not have, therefore,
internal shocks. Realistic engines are likely characterized by
intrinsic variability, and a better agreement with the Amati and
Yonetoku ensemble correlations could be obtained from a light curve
with the same peak frequency and peak luminosity but with periods of
very weak emission.

Additional physics might also be required to obtain a better match of
the synthetic spectra with observations. Shocks likely inject
non-thermal particles that are neglected by MCRaT and by the
\cite{Ito15} code alike. Magnetic fields and the possibility of
injection of soft photons from synchrotron are also ignored
\citep{Vurm11,Vurm16}.  Additional dissipation is also provided by
magnetic reconnection \citep{Giannios07,McKinney02}. While adding such
effects would not be trivial, breaking the embarrassingly parallel
structure of the code and making the whole calculation longer, it is
not impossible and we are currently developing a version of the MCRaT
code that takes into account non-thermal particles and synchrotron
emission.  Since all the missing effects are likely to increase the
non-thermal aspect of the spectrum, our the results presented here can
be seen as a lower limit on the non-thermal features of the
photospheric spectrum from GRB jets.

Finally, we would like to comment on the fact that in no case we found
the need to add a low-temperature thermal component to  the
  spectral model used to fit our spectra. Yet, such a component has
been detected in several bursts and is considered one of the proofs of
the existence of a photospheric contribution to the GRB prompt
spectrum
\citep{Guiriec11,Ghirlanda13,Guiriec13,Burgess14,Nappo16}. Our
results, as well as Ito's results seem to show that the appearance of
the photospheric component is always non-thermal and at significantly
higher frequencies than those at which the thermal components are
detected (typically a few keV to a few tens of keV). The thermal peaks
detected in BATSE and Fermi observations are therefore either arising
from a different component of the outflow that is not present in the
simulations or are a signature of a weak photosphere in a jet
dominated by non-thermal and/or non-baryonic energy, such as in a
Poynting dominated outflow.

\acknowledgments I would like to thank the anonymous referee for
constructive comments that improved the clarity of this paper. I would
like to thank Hirotaka Ito and Hiro Nagataki for useful discussions
and for sharing the results of their code with me prior to
publication. I also would like to thank Andrei Beloborodov, Riccardo
Ciolfi, Bruno Giacomazzo, Dimitrios Giannios, Diego L\'opez-C\'amara,
and Brian Morsony, for useful comments and suggestions. This work was
supported in part by NASA Swift GI grant NNX15AG96G.

\appendix

\section{Derivation of the photon weight}

The luminosity of a uniform jet can be written as (e.g., Morsony et
al. 2010):
\begin{equation}
L_j=\Omega_j R^2 \left(4p+\rho' c^2\right)\Gamma^2  c
\end{equation}
where $p$ is the pressure of the jet material, and $\rho'$ its
comoving density. For a jet dominated by radiation ($p=aT'^4/3$, and
$p\gg \rho' c^2$), and considering the solid angle of a one-sided jet,
the equation simplifies to:
\begin{equation}
L_j=\frac{8\pi}{3}\left(1-\cos\theta_j\right) R^2 aT'^4\Gamma^2  c
\end{equation}

Deriving with respect to the polar angle we find the luminosity per
unit polar angle (the simulated quantity in cylindrical symmetry) and
dividing by the average photon energy $h\bar\nu=\Gamma h\bar\nu'$
we obtain the number of photons per unit polar angle as:
\begin{equation}
\frac{dN_\gamma}{dt \,d\theta}=\frac{1}{h\bar\nu}\frac{dL}{d\theta}
=\frac{8\pi}{3}\sin\theta_jR^2 \frac{aT'^4}{h\bar\nu'}\Gamma  c
=\frac{8\pi}{3}\sin\theta_jR^2 n_\gamma'\Gamma  c
=\frac{8\pi}{3}\sin\theta_jR^2 \xi T'^3 \Gamma  c
\end{equation}
This should be equal to the number of photons we inject in the
simulation times the weight assigned to each photon:
\begin{equation}
\frac{dN_\gamma}{dt \,d\theta}=\frac{w_{\gamma,i}N_{\rm{inj}}}{\delta t\,\delta\theta}
\end{equation}

Solving for $w_{\gamma,i}$ yields Eq.~\ref{eq:weight}.

\end{document}